# Nanometric voids as optical antennas for rewritable momentum-engineered photonics in silicon

*Aleksei I. Noskov[1], Evan P. Garcia[1], Haiping Sun[2], Eric O. Potma[1], Dmitry A. Fishman[1]*

[1]*Department of Chemistry, University of California, Irvine, Irvine, CA 92697, USA*

[2]*Irvine Material Research Institute, University of California, Irvine, Irvine, CA 92697, USA*

## Abstract

Optical antennas are widely used to localize electromagnetic fields far below the diffraction limit, enabling enhanced light–matter interactions across nanophotonics. Yet the regime in which optical confinement approaches the electronic de Broglie wavelength in a solid - where the photon momentum distribution broadens sufficiently to relax optical selection rules - remains largely unexplored. Here we show that nanometric voids embedded within crystalline silicon act as such optical antennas, dramatically altering the optical response of an indirect semiconductor without the introduction of any foreign material. Using an electrically induced melt–quench process, we generate nanometric voids throughout bulk silicon, confirmed by high-resolution electron microscopy, diffraction analysis, Fourier-filtered lattice reconstruction, elemental mapping, and supported by optical and vibrational spectroscopies. The void-containing silicon exhibits intense broadband photo- and electroluminescence spectrally indistinguishable from that produced by metallic or semiconductor nanoconfiners of similar dimensions, establishing that dielectric discontinuity, not confiner composition, governs the observed momentum-assisted optical transitions. The luminescence can be repeatedly written, erased, and rewritten through alternating electrical conditioning and optical recrystallization. These findings establish nanometric voids as a previously unexplored platform for extreme optical confinement and demonstrate that photonic functionality can be embedded and reconfigured directly within bulk silicon.

Optical antennas are structures that convert freely propagating optical radiation into localized electromagnetic energy, or vice versa, thereby enhancing and controlling light–matter interactions on length scales far below the wavelength of light. Inspired by their radio-frequency counterparts, optical antennas have become central elements of modern nanophotonics, enabling electromagnetic confinement beyond the diffraction limit through plasmonic, dielectric, and hybrid nanostructures.[1-5] Their ability to increase the local optical density of states and generate strong field gradients underpins a broad range of phenomena, including enhanced spectroscopy, nonlinear optical processes, nanoscale imaging, and photodetection.[6-12] Yet the regime in which optical confinement approaches nanometric dimensions remains comparatively unexplored. Beyond field enhancement via the Purcell effect[13], extreme confinement has been predicted to enable interactions beyond the dipole approximation, including higher-order multipolar transitions and momentum-restricted electronic transitions that are otherwise inaccessible in conventional optics.[2,14,15]

At these short length scales, the optical field can no longer be viewed as a propagating wave characterized by its free-space momentum.[2,16-19] Instead, strong spatial confinement produces a broad distribution of optical momenta, whose bandwidth is determined by the confinement length scale and can extend the accessible photon momentum into the range of electronic wavevectors in solids. In turn, this relaxes conventional momentum-selection rules and enables optical transitions that are inaccessible under far-field illumination.[16-18,20]

Indirect semiconductors provide a particularly sensitive platform for exploring this regime. In silicon, the conduction-band minimum and valence-band maximum are misaligned in momentum space. This misalignment restricts optical transitions across the visible and near-infrared range, requiring phonon assistance from the crystal lattice. [21-23] As a result, optical absorption near the band edge is weak and radiative recombination is intrinsically inefficient.

Because electron momentum conservation constitutes the dominant limitation to efficient optical transitions, even modest changes in the accessible optical momentum can profoundly alter the optical response. Recently, it has been demonstrated that ultrasmall metallic nanoconfiners dramatically enhance optical absorption[24-26] and produce intense broadband luminescence[27] from bulk silicon. These studies revealed that confinement approaching only a few nanometers fundamentally alters the optical response of an indirect semiconductor. Replacing metallic confiners with ultrasmall silicon nanocrystallites below 1.5 nm in size yielded spectrally indistinguishable optical responses in both frequency and time domains, despite the absence of common electronic, plasmonic, or chemical characteristics.[28] These observations suggested that the governing parameter was not the composition of the confiner, but rather the extreme optical confinement associated with a nanometric dielectric discontinuity.

This conclusion leads to a fundamental prediction. *If momentum-engineered optical transitions arise from confinement rather than composition, then any nanometric dielectric discontinuity capable of localizing optical fields should generate a similar response*. Nanometric voids would provide perhaps the most stringent test of this hypothesis. Unlike metallic or semiconductor nanoparticles, voids introduce no foreign materials, no plasmonic resonances, and no additional electronic states above the bandgap. Their only role is to establish a nanometric region of strongly reduced dielectric function embedded within the host crystal.

Here we demonstrate that voids generated within bulk silicon via an electrically induced melt–quench process dramatically alter the optical response of the host material. Structural characterization reveals a dense population of nanometric low-density regions consistent with void formation, while optical and vibrational spectroscopies provide independent evidence of the associated lattice reconstruction. The void-containing material produces broadband photo- and electroluminescence spectrally indistinguishable from that observed using metallic and semiconductor nanoconfiners, while remaining fully rewritable through alternating electrical conditioning and light-induced recrystallization. Together, the results establish nanometric dielectric discontinuities as a material-independent route to momentum-assisted optical transitions and demonstrate a rewritable photonic state embedded directly within bulk silicon.

## Results

### Electrical conditioning creates a luminescent silicon state

To create a model system in which nanometric voids serve as optical antennas within bulk silicon, we use a commercial silicon p-i–n photodiode as the electrical platform (see *Methods*). Under significant forward bias, holes driven into the top p-type layer recombine with trapped electrons at dopant sites through Shockley–Read–Hall nonradiative processes, releasing energy locally as heat. [29,30] The collective action of these recombination sites gradually warms the bulk material to measurable temperatures, as confirmed by Raman thermometry (Supplementary Information Part I, Figures SF1-SF3). Although the average lattice temperature remains well below the melting point of silicon (Figures SF2-SF4), individual recombination sites are expected to experience substantially higher transient temperatures, sufficient to produce a locally molten phase. This local melting is expected to be accompanied by thermal expansion of the lattice at the recombination site, generating the transient low-density regions. When the applied forward bias is abruptly removed, the surrounding crystalline lattice acts as a rapid heat sink, quenching these regions before they can recrystallize, leaving behind a dense population of nanometric voids embedded within an otherwise crystalline silicon matrix (Figure 1a). Gradually reducing the forward bias produces no observable effect, confirming that rapid quenching, rather than prolonged heating, governs the formation and preservation of the void (Figure SF4). Electrical conditioning leaves the current-voltage characteristics unchanged, confirming that the melt–quench process does not compromise device functionality. (Supplementary Information Part II, Figures SF6 and SF7).

From this point onward, we focus on the optically accessible top silicon layer of the photodiode. The pristine bulk semiconductor exhibits only weak intrinsic silicon photoluminescence associated with phonon-assisted carrier recombination from the conduction-band minimum near 1100 nm under 532 nm optical excitation (Figure 1b, grey). Bulk amorphous silicon, examined as a reference to rule out structural disorder as the origin of the emission, exhibits only the characteristic broad Raman band of the disordered matrix near 480 $cm^{-1}$, with no detectable broadband photoluminescence (Figure 1b, purple). In contrast, electrically conditioned bulk silicon reveals intense broadband photoluminescence spanning the visible to near-infrared spectral range (Figure 1b, orange). The emission is observed under both 532 nm and 785 nm excitation, confirming its photoluminescent origin rather than electronic Raman scattering, surface contamination, multiphoton-excited emission, or other excitation-wavelength-dependent artifacts.

Notably, the photoluminescence decay is characterized by a lifetime of approximately 2 ns, a value at odds with any known radiative mechanism in native silicon. Phonon-assisted recombination from the conduction-band minimum, the dominant radiative pathway in bulk silicon, proceeds on timescales of tens of microseconds.[21] Silicon quantum dots emitting in the visible and near-infrared exhibit radiative lifetimes in the range of 1-100 μs, a direct consequence of the phonon-mediated nature of recombination even under quantum confinement.[31-35] Known structural defect centers in silicon are equally incompatible: W centers emit at a single narrow line at 1218 nm and quench entirely above 80 K, while dislocation-related D-lines produce narrow deep sub-bandgap emission at 1300-1600 nm [36]. Both exhibit radiative lifetimes of tens to hundreds of nanoseconds, orders of magnitude slower than the observed ~2 ns decay, and neither produces broadband visible emission at room temperature.[37,38]

The spectral response of the void-containing silicon is indistinguishable from that previously observed in silicon coupled to 1-2 nm sized metallic and semiconductor nanoconfiners, both in the frequency domain (Figure 1d) and in the time domain (Figure 1e). The complete agreement among these structurally distinct systems is particularly remarkable because the confiners differ fundamentally in composition, electronic structure, and optical properties. The only common features are the presence of bulk silicon and nanometric light confiners in proximity to it. This observation strongly suggests that the emission (a) originates from the bulk semiconductor host and (b) is governed not by the specific chemical identity of the confiner, but by the presence of nanometric dielectric discontinuities capable of producing extreme optical confinement.

If the observed optical response is indeed associated with nanometric dielectric discontinuities created during electrical conditioning, then repairing the underlying lattice structure should eliminate the effect. To test this hypothesis, we locally irradiated electrically conditioned silicon with modest 25 mW continuous-wave laser radiation at 532 nm, writing a circular pattern across the surface (Figure 1f and 1g). Within the irradiated regions, the heat absorbed recrystallizes the lattice, erasing the bright broadband luminescence and restoring the weak intrinsic photoluminescence characteristic of pristine silicon, while the surrounding areas retain their intense luminescent properties (Figure 1g). The ability to create and erase the luminescent state further indicates that the effect originates from a reversible structural modification of the host lattice rather than from permanent chemical transformations.

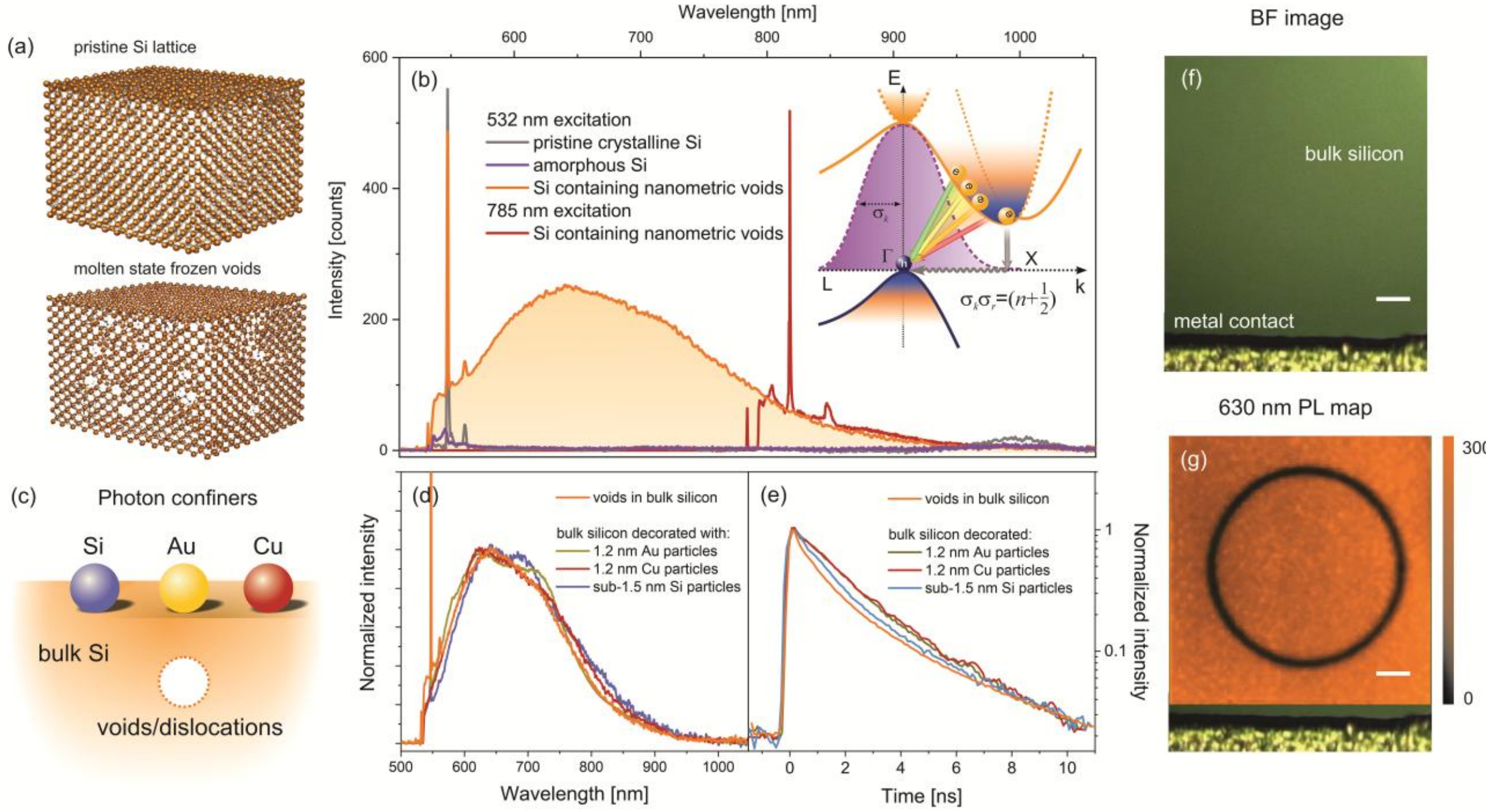

**Figure 1. Momentum-engineered photonics states in silicon via void-assisted confinement.** (a) Atomic models of pristine silicon (top) and electrically conditioned silicon containing frozen molten-state nanometric voids (bottom). (b) Photoluminescence spectra of pristine (grey) and electrically conditioned silicon under 532 nm (orange) and 785 nm (red) excitation. The void-containing silicon exhibits intense broadband emission spanning the visible and near-infrared, observed under both excitation wavelengths, confirming the photoluminescent origin of the observed emission. Inset: schematic band diagram illustrating voids acting as optical antennas that expand the photon momentum distribution $\sigma_k$, enabling diagonal transitions across the Brillouin zone. (c) Schematic representation of bulk silicon decorated with optical field confiners of different origin. Normalized emission spectra in the frequency (d) and time (e) domains comparing void-containing silicon emission with silicon decorated with metallic (Au, Cu) and semiconductor (Si) nanoconfiners of sub-2 nm dimensions. The signals are indistinguishable across all confiner types. (f) Bright-field optical image of the void-containing silicon surface near the metal contact. Scale bar, 20 µm. (g) Confocal photoluminescence map at 630 nm of the same region following focused continuous-wave laser irradiation. The dark circle corresponds to a locally recrystallized region, where the broadband emission is fully suppressed, while the surrounding conditioned silicon retains its intense photoluminescence. Scale bar 20 µm.

**Visualization of nanometric voids in crystalline matrix**

Visualization of nanometric voids within a crystalline matrix is challenging because their dimensions approach only a few lattice constants and their density contrast is small. To establish the presence of voids, we combined complementary imaging, diffraction, elemental mapping, and image-reconstruction techniques. To determine the structural origin of the luminescent state, cross-sectional lamellae were prepared from pristine and electrically conditioned silicon by focused ion beam (FIB) milling and examined by scanning transmission electron microscopy (STEM) (Figure 2).

In contrast to pristine silicon, which exhibits featureless, homogeneous contrast throughout the bulk (Figure 2e,f), the conditioned samples contain a dense population of sub-3 nm features distributed throughout the silicon volume (Figure 2c,d, see also Supplementary Information Part III and Figure SF7). These features appear dark in high-angle annular dark-field imaging and bright in bright-field STEM - *a contrast reversal that is consistent with a locally reduced mass density and is inconsistent with implantation, surface deposition, or heavier elemental inclusions*.

Elemental mapping further supports this interpretation. Energy-dispersive X-ray spectroscopy (EDS) detects no foreign elements within the bulk silicon cross-section, including oxygen, gallium, or platinum potentially introduced during FIB preparation and lamella handling (Supplementary Information Part IV, Figure SF10). Instead, the silicon elemental map itself exhibits intensity minima that spatially correlate with the dark-field and bright-field STEM contrast (Figure 2g), confirming that the features represent regions of reduced silicon density rather than chemically distinct inclusions. The void density estimated from both STEM and EDS silicon maps is on the order of $10^{17}$-$10^{18}$ $cm^{-3}$, comparable to typical dopant concentrations in the p-layer and consistent with nucleation at dopant-associated recombination sites during the melt–quench process. Size distribution analysis of bright-field STEM contrast features yields a void diameter distribution peaked at 1.3-1.5 nm, with the vast majority of voids falling below 3 nm (Supplementary Information Part III, Figures SF8 and SF9).

Selected-area electron diffraction (SAED) provides independent evidence for the structural transformation of the electrically conditioned silicon. Pristine silicon produces sharp diffraction spots characteristic of a well-ordered diamond cubic lattice (Figure 3a). In contrast, electrically conditioned silicon exhibits crystalline reflections accompanied by pronounced diffuse scattering and signatures of lattice disorder (Figure 3b), indicating coexistence of crystalline silicon with amorphous or strongly distorted local regions.

Fourier-filtered lattice reconstruction, obtained by selecting crystalline spatial frequencies in the HR-TEM FFT and applying an inverse FFT, yields a map of the periodic lattice (see also Supplementary Information Part IV, Figure SF11). Applying this procedure to pristine silicon yields a continuous, uninterrupted lattice, confirming the integrity of the crystalline matrix (Figure 3c,d). The same analysis applied to electrically conditioned silicon reveals a crystalline backbone interrupted by nanometric regions of suppressed lattice contrast (Figures 3e and 3f). *These interruptions corroborate the low-density regions identified by STEM and reveal nanometric* *disruptions of the crystalline lattice consistent with embedded voids, providing atomic-scale evidence for the void-containing phase.* Size analysis of the suppressed-contrast regions yields a void diameter distribution peaked at 1.0-1.2 nm, consistent with the bright-field STEM size distributions.

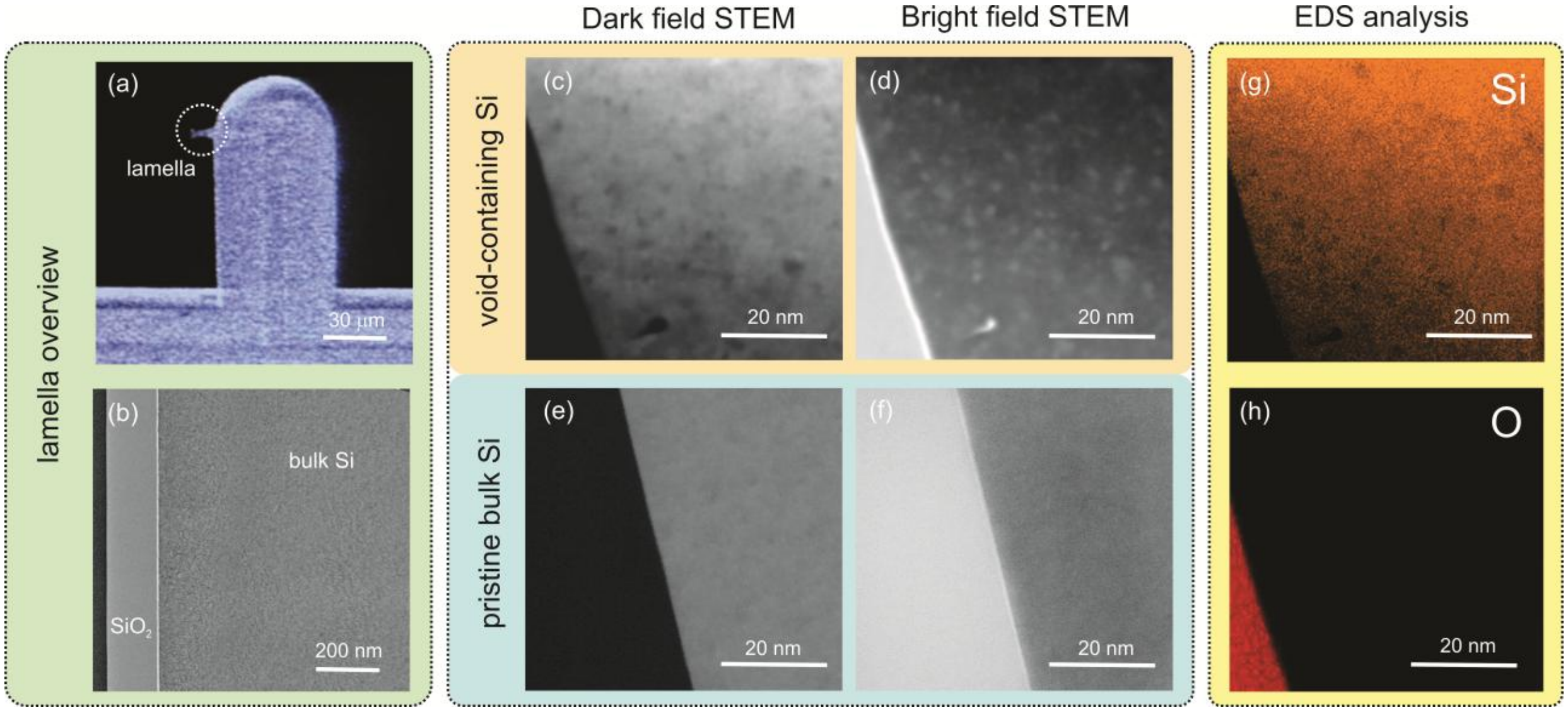

**Figure 2. Direct visualization of nanometric voids in silicon.** (a) Low-magnification SEM overview of the FIB-prepared lamella (in dashed circle) mounted on the TEM holder. Scale bar 30 µm. (b) Cross-sectional STEM image of the lamella, showing the bulk silicon layer capped by the passivation $SiO_2$ layer on the left. Scale bar 200 nm. (c, d) Dark-field (DF) and bright-field (BF) STEM images of electrically conditioned silicon, respectively. A dense population of sub-3 nm features is visible throughout the silicon volume, appearing dark in DF (c) and bright in BF-STEM (d) - a contrast reversal consistent with locally reduced silicon density. Size analysis of the bright-field contrast features yields a void diameter distribution peaked at 1.3-1.5 nm (Supplementary Information Part III, Figure SF8 and SF9). Scale bars 20 nm. (e, f) DF and BF-STEM images of pristine bulk silicon under identical imaging conditions, showing featureless, homogeneous contrast with no evidence of nanometric features. Scale bars 20 nm. (g) EDS elemental map of silicon acquired from the conditioned sample, revealing intensity minima that spatially correlate with the DF and BF-STEM contrast, confirming that the sub-3 nm features correspond to regions of reduced silicon density. Scale bar 20 nm. (h) EDS elemental map of oxygen, showing oxygen localized exclusively within the SiO2 surface passivation layer with no detectable oxygen within the bulk silicon volume (see also Supplementary Information Part III and Figure SF7). Scale bar 20 nm.

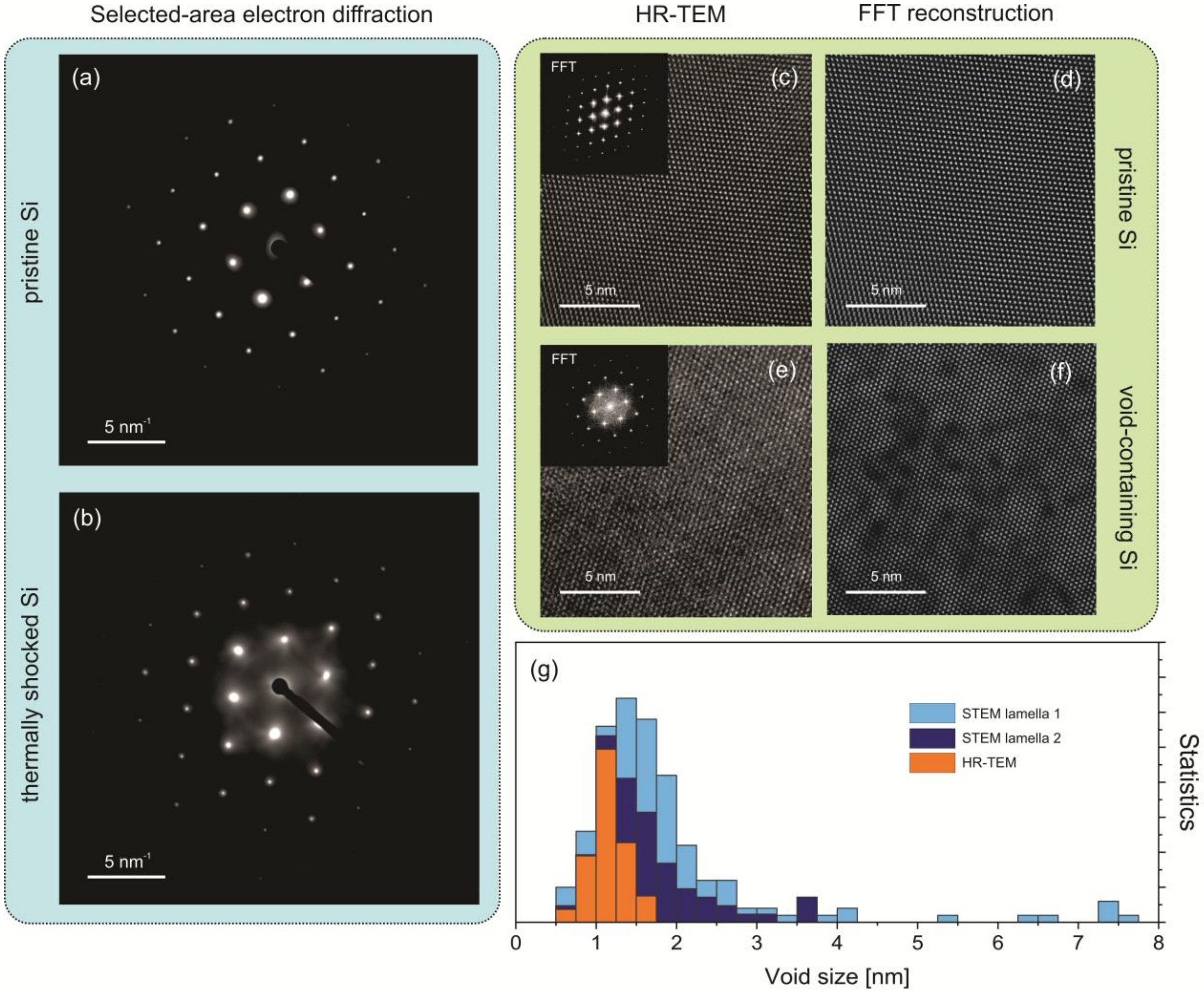

**Figure 3. Selected-area electron diffraction and Fourier-filtered lattice reconstruction reveal nanometric voids in electrically conditioned silicon.** (a, b) Selected-area electron diffraction (SAED) patterns of pristine (a) and electrically conditioned (b) silicon. Pristine silicon displays sharp Bragg spots consistent with a well-ordered diamond cubic lattice. Conditioned silicon retains crystalline reflections, but exhibits additional diffuse scattering and broadened spot profiles, indicating the coexistence of crystalline silicon with amorphous or strongly distorted local regions. Scale bars 5 $nm^{-1}$. (c) HR-TEM image of pristine bulk silicon with FFT inset showing sharp, well-defined diffraction spots. Scale bar 5 nm. (d) Fourier-filtered lattice reconstruction of (c), revealing a continuous, uninterrupted periodic lattice confirming the integrity of the crystalline matrix. Scale bar 5 nm. (e) HR-TEM image of electrically conditioned silicon with FFT inset showing diffuse and broadened diffraction features. Scale bar 5 nm. (f) Fourier-filtered lattice reconstruction of (e), revealing a crystalline backbone interrupted by nanometric dark regions of suppressed lattice contrast, directly visualizing nanometric voids embedded within the crystalline silicon matrix. Size analysis of the suppressed-contrast regions yields a void diameter distribution peaked at 1.0-1.2 nm. Scale bar 5 nm. (g) Overlay of void size distributions obtained from STEM and HR-TEM electron microscopy modalities: two lamellae bright-field STEM datasets (light blue - Figure 2d; dark blue - Supplementary Figure SF5) and Fourier-filtered HR-TEM lattice reconstruction (orange - Figure 3f). All

three distributions are peaked in the 1.0–1.5 nm range, demonstrating excellent agreement across independent imaging modalities and independently prepared specimens.

## Spectroscopic fingerprints of void formation

The formation of nanometric voids necessarily removes silicon atoms from their original lattice positions, requiring substantial local lattice reconstruction. The displaced atoms must be incorporated into the surrounding lattice, forming stable point-defect complexes and extended defect structures. Among the most well-established signatures of such reconstruction are W centers, compact tri-interstitial silicon clusters with trigonal symmetry[37,39] and dislocation-related D-line defects associated with local lattice distortion and strain[36]. Low-temperature near-IR photoluminescence spectroscopy, where these defects are known to be spectrally active, therefore provides an independent spectroscopic probe of the structural modifications accompanying void formation.

As shown in Figure 4b, pristine silicon exhibits only the conventional phonon-assisted band-edge emission characteristic of indirect recombination near 1100 nm at 80 K. In electrically conditioned silicon, this emission is strongly depleted, while sharp spectroscopic features associated with W-center zero-phonon lines and dislocation-related D-lines emerge simultaneously. This depletion mirrors the suppression of band-edge luminescence observed at room temperature (Figure 1b) and is consistent with observations in silicon coupled to metallic nanoconfiners, where momentum-engineered optical transitions were likewise accompanied by a suppression of the conventional phonon-assisted pathway [27]. The simultaneous appearance of W-center and D-line signatures provides direct spectroscopic evidence of lattice reconstruction associated with void formation.

These defect signatures serve as spectroscopic fingerprints of void formation rather than the origin of the broadband optical response. W centers and dislocation-related D-lines produce narrow near-infrared emission that is absent above 80 K and exhibit radiative lifetimes ranging from tens to hundreds of nanoseconds - far longer than the observed ~2 ns decay. [37,38] Neither defect family generates broadband visible-to-near-infrared emission. The defect signatures therefore independently confirm void-associated lattice reconstruction, while the broadband room-temperature emission arises from photon momentum-assisted optical transitions in the surrounding bulk silicon enabled by the nanometric dielectric discontinuities.

Additional evidence for a structurally distinct low-density phase is provided by vibrational spectroscopy. Electrically conditioned silicon exhibits a pronounced Boson peak centered near 170 $cm^{-1}$ - a spectral feature characteristic of low-density amorphous material absent in pristine crystalline silicon (Figure 4c).[40,41] The emergence of this mode provides independent vibrational evidence that electrical conditioning generates a structurally distinct low-density phase beyond simple point-defect formation, consistent with the melt–quench picture in which locally molten regions solidify into low-density amorphous domains.

The Boson peak position is related to the structural correlation length ξ of the disordered phase through the Ioffe–Regel criterion, $\xi \approx \upsilon_T/(c\upsilon_{BP})$, where $\upsilon_T$ is the transverse acoustic velocity, $\upsilon_{BP}$ is the Boson peak frequency, and $c$ is the speed of light in vacuum.[40] With estimated $\upsilon_T$=6500 m/s [42] and $\upsilon_{BP}$=170 cm$^{-1}$, a characteristic correlation length of approximately 1.3 nm is found, reflecting the size scale of the low-density reconstructed regions associated with void formation. This length scale corroborates the void dimensions inferred from STEM and HR-TEM imaging and the sub-2 nm confinement regime required for momentum-engineered optical transitions, providing convergence of independent structural and vibrational measurements on the same nanometric length scale (Supplementary Information Part VI, Table ST1).

Spatially resolved Raman and photoluminescence imaging further confirm the connection between structural and optical signatures. Confocal maps of the Boson peak intensity at 170 cm$^{-1}$ and broadband photoluminescence are in excellent spatial agreement (Figure 4d,e), with both signals suppressed within the laser-recrystallized region and retained in the surrounding void-containing silicon. This spatial correlation provides an important bridge between the structural characterization in Figures 2 and 3 and the optical phenomena shown in Figure 1, directly linking the low-density nanometric amorphous phase to the momentum-engineered optical response.

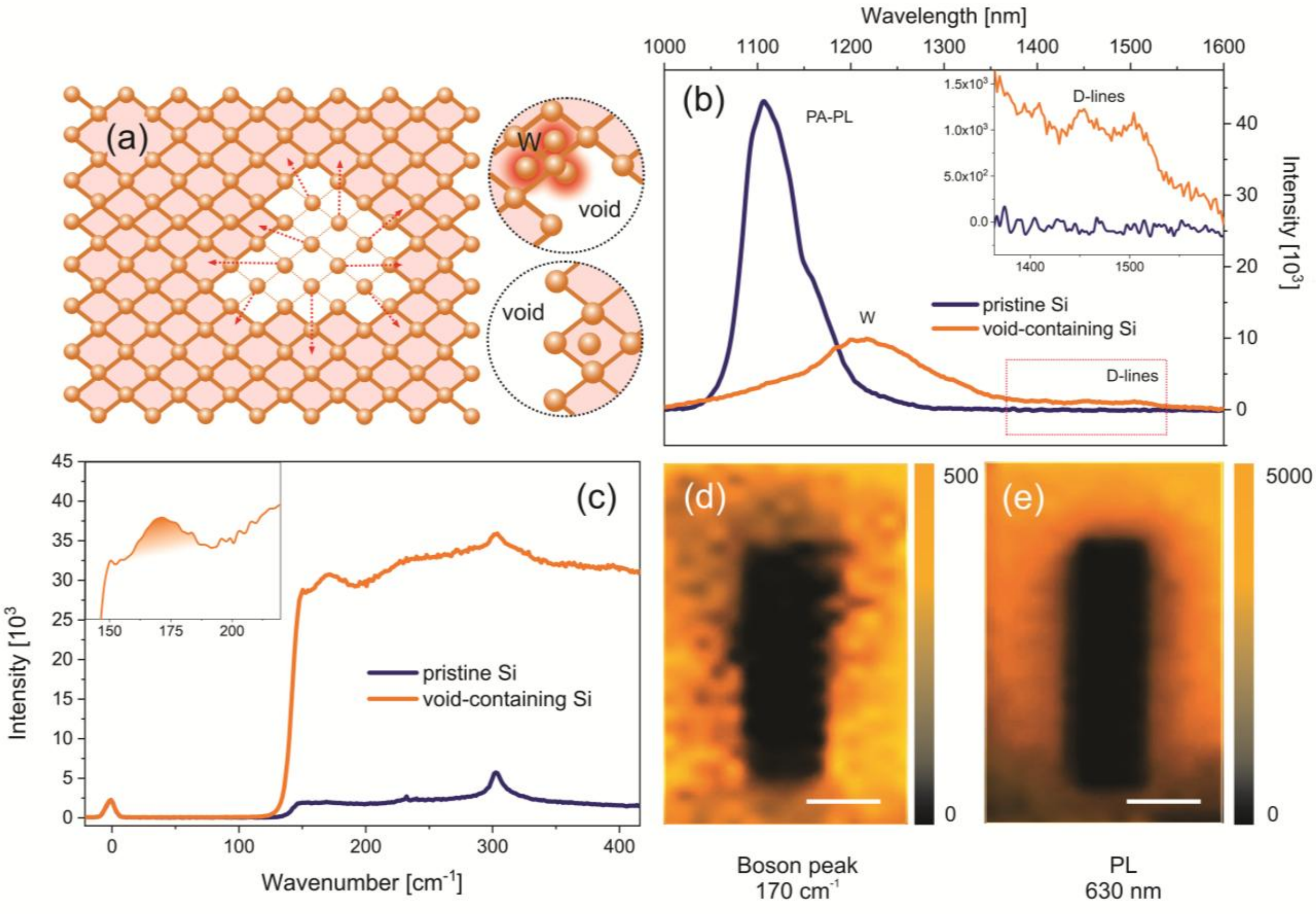

**Figure 4. Spectroscopic fingerprints of void formation.** a) Schematic of the electrically conditioned silicon lattice showing a nanometric void formed by displacement of silicon atoms from their original lattice sites. Red dashed arrows indicate the directions of atomic displacement. Insets: the displaced atoms migrate

to neighboring unit cells and may form a stable tri-interstitial W center. (b) Low-temperature (80 K) photoluminescence spectra of pristine (blue) and electrically conditioned (orange) silicon in the near-infrared. Pristine silicon exhibits only the conventional phonon-assisted band-edge emission near 1100 nm. In conditioned silicon, this emission is strongly depleted while sharp W-center zero-phonon lines appear simultaneously. Inset: expanded spectral region showing dislocation-related D-lines present in conditioned silicon and absent in pristine silicon. (c) Raman spectra of pristine (blue) and electrically conditioned (orange) silicon. Conditioned silicon exhibits a pronounced Boson peak at ~170 $cm^{-1}$, characteristic of a low-density amorphous phase, and entirely absent in pristine crystalline silicon. The Boson peak position corresponds to a structural correlation length of ~1.3 nm, consistent with the void dimensions identified by STEM. (d) Confocal Raman map of Boson peak intensity at 170 cm-1. The dark rectangular region corresponds to a locally laser-recrystallized area in which the amorphous void-containing phase has been eliminated. Scale bar 20 μm. (e) Confocal photoluminescence map at 630 nm acquired from the same region as (d). Scale bar 20 μm.

**Reversibility and rewritable optical antennas**

The optical writing experiment introduced in Figure 1 demonstrated that laser recrystallization locally suppresses the void-induced photoluminescence. Here we show that the photonic state can be repeatedly created and erased within the same bulk semiconductor. Electrically conditioned silicon initially exhibits spatially homogeneous broadband photoluminescence (Figure 5c). Focused continuous-wave laser irradiation can locally recrystallize the material, erasing the luminescent state along the written path while leaving the surrounding conditioned silicon unchanged (Figure 5d).

Following optical erasure of the circular pattern, a second electrical conditioning step reapplies the void-containing luminescent state across the entire surface, after which a new pattern of crosses can be written by local laser recrystallization (Figure 5e). The ability to repeatedly create, erase, and regenerate these patterns demonstrates that the photonic state is rewritable and governed by reversible structural modification of the silicon lattice rather than irreversible contamination, permanent damage, or extrinsic emitters. The recovery of broadband emission following reconditioning (Figure 5a, Supplementary Figure SF13) confirms regeneration of the momentum-engineered photonic state.

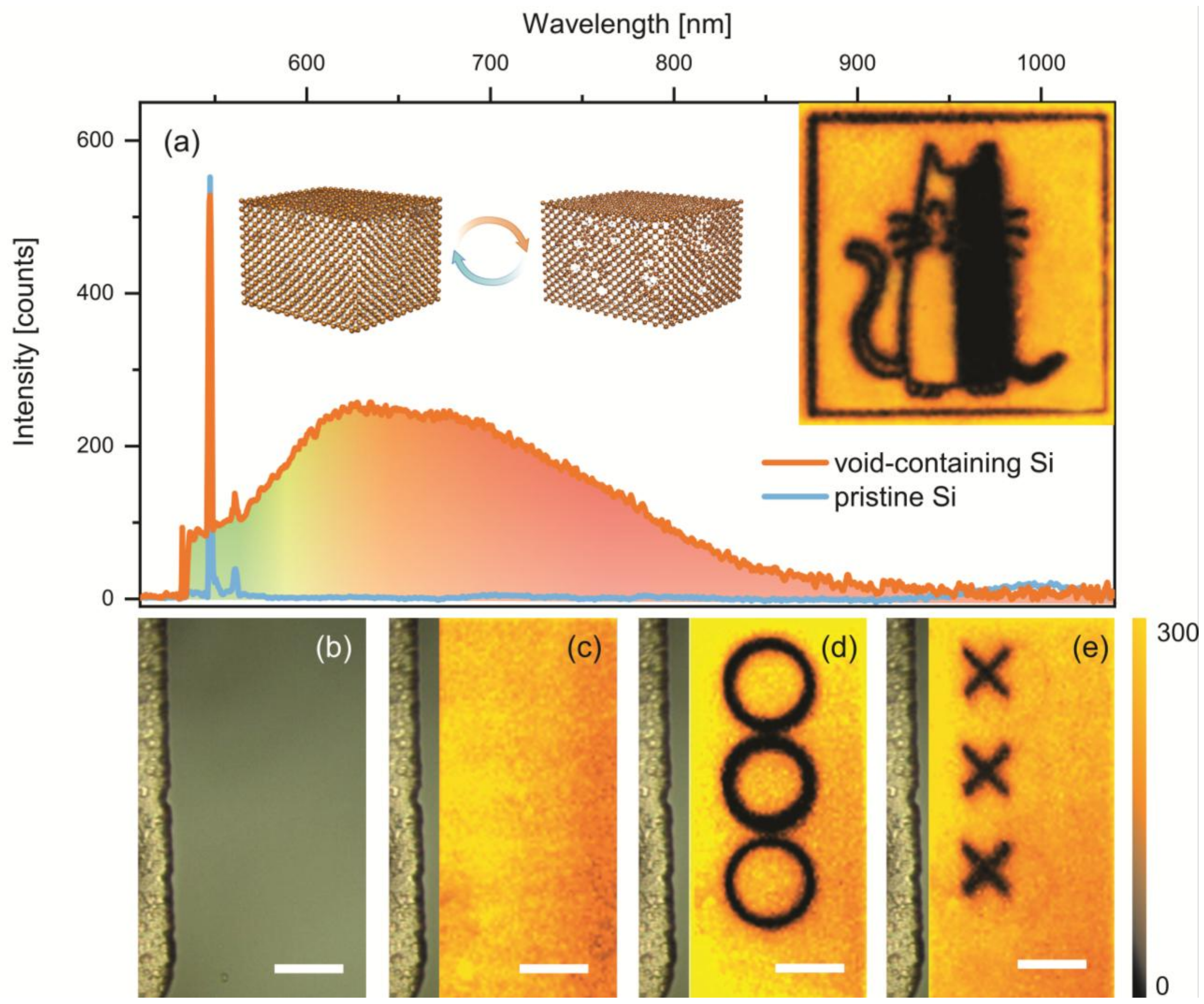

**Figure 5. Reversible write–erase cycling of momentum-engineered photonic states in bulk silicon.** (a) Photoluminescence spectra of electrically conditioned (orange) and laser-recrystallized (blue) silicon, demonstrating complete spectral reversibility between the void-containing luminescent state and the restored pristine state. Inset schematics: atomic models of pristine (left) and void-containing (right) silicon lattices connected by a reversibility arrow, illustrating the write–erase cycle. Inset image: photoluminescence map of a "Schrödinger cat in a box" pattern written by laser recrystallization into electrically conditioned silicon, demonstrating arbitrary-pattern optical writing capability (see also Figure SF14). (b) Bright-field optical image of the electrically conditioned silicon surface near the metal contact. Scale bar 20 μm. (c) Confocal photoluminescence map of the uniformly conditioned silicon surface, showing homogeneous broadband emission across the entire region. Scale bar 20 μm. (d) Photoluminescence map after focused continuous-wave laser irradiation, showing two circular patterns written into the luminescent surface. Scale bar 20 μm. (e) Photoluminescence map following a second electrical conditioning cycle and subsequent laser writing, showing a new pattern of crosses written on the same surface. The complete recovery of uniform emission before writing confirms full regeneration of the void-containing luminescent state. Scale bar 20 μm.

**Electroluminescence from void-containing silicon**

The same void-containing silicon also supports electrically driven emission. Figure 6 demonstrates electroluminescence from a device comprising a void-containing p-type silicon slab with two surface electrodes separated by approximately 1.5 mm (Figure 6a). Under a 7 V bias, electrically conditioned devices exhibit bright electroluminescence visible to the naked eye from the silicon surface in proximity to the electron-injecting electrode (Supplementary Video 1), whereas pristine devices show no detectable emission under comparable conditions (Figure 6b,c). Only under substantially higher voltage (30 V) does pristine silicon exhibit electroluminescence, arising from phonon-assisted recombination at the conduction-band minimum - a fundamentally different emission mechanism operating at more than four times the voltage.

The lateral electrode geometry and modest applied voltage conclusively rule out conventional electroluminescence mechanisms (see Supplementary Information Part VIII for the detailed discussion and analysis). The average electric field across the 1.5 mm electrode separation at 7 V is 47 V/cm - four to five orders of magnitude below the ~$3 \times 10^5$ V/cm required for impact ionization and avalanche electroluminescence in silicon.[43] Both electrodes contact the p-type surface layer, and current flows laterally through the void-containing region without involving the buried p-n junction or its depletion region, ruling out junction-based electroluminescence. Emission is observed near the needle electrode, consistent with locally enhanced carrier recombination in the void-containing phase. The absence of emission from pristine silicon under identical conditions confirms that the void-containing phase, rather than the applied field geometry, is the critical requirement for electroluminescence.

Together, these results demonstrate that nanometric dielectric discontinuities embedded within bulk silicon activate both optically and electrically driven emission pathways. The emergence of electroluminescence from bulk silicon containing no foreign light-emitting material provides a direct device-level manifestation of momentum-engineered photonics in silicon and establishes that the confinement-induced radiative pathways remain accessible under electrical carrier injection.

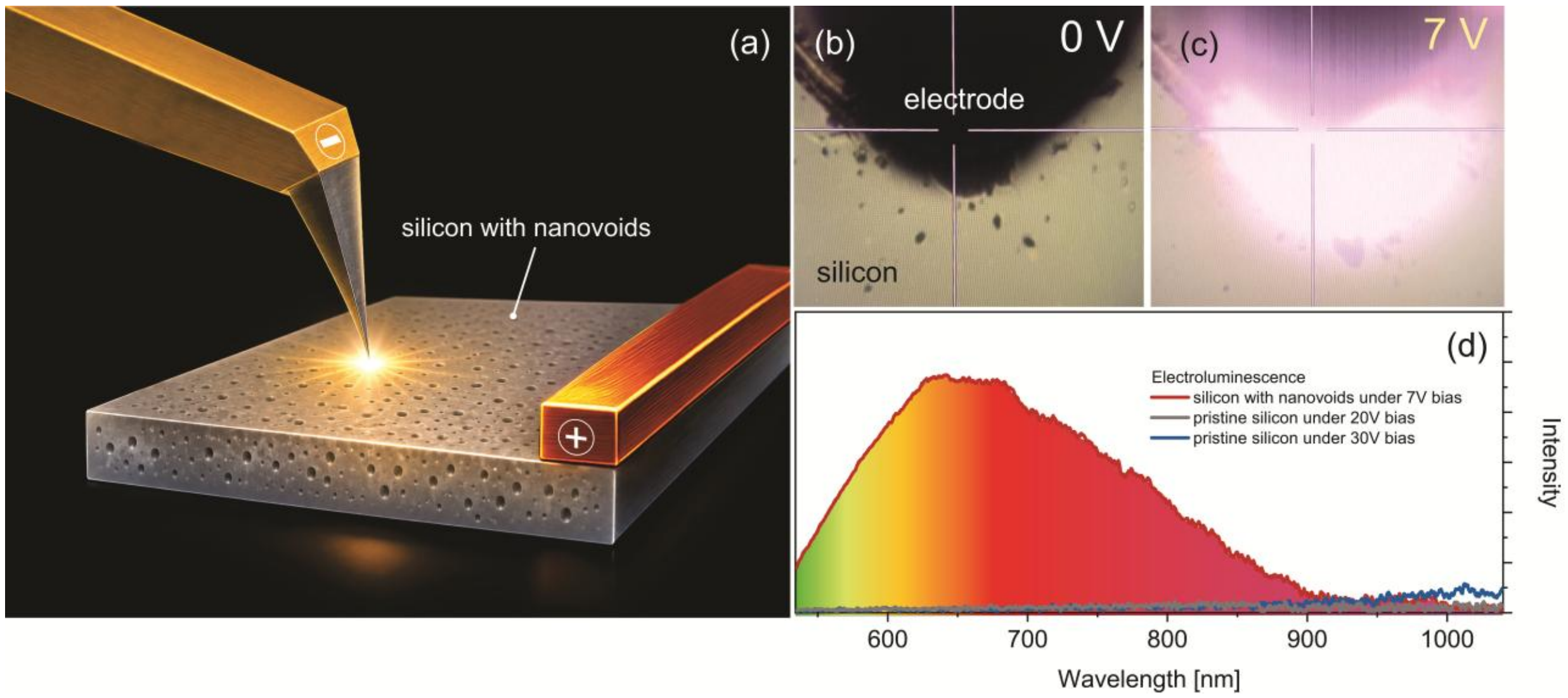

**Figure 6. Injection electroluminescence from void-containing bulk silicon.** (a) Schematic illustration of the electroluminescence measurement geometry. A probe serving as the electron-injecting electrode contacts the surface of a void-containing silicon p-doped slab, with a second electrode at a 1.5 mm distance providing the positive contact. Emission is observed from the silicon surface close to the injection point. (b, c) Still frames of Supplementary Video 1, a video recording of the void-containing silicon device under 0 V (b) and 7 V (c) external injection voltage. (d) Electroluminescence spectra of electrically conditioned, void-containing silicon (red) and pristine silicon under 20 V (grey) and 30 V (blue). The void-containing device produces intense broadband emission spanning the visible and near-infrared, while pristine silicon shows no detectable electroluminescence.

## Discussion

The central result of this work is that nanometric voids reproduce the optical response previously observed with metallic and semiconductor nanoconfiners with remarkable fidelity. This convergence across structurally and chemically distinct systems points to a single governing parameter: the nanometric dielectric discontinuity within the host material.

This conclusion is most naturally understood within the framework of optical antennas. Nanometric voids offer a uniquely minimal realization of this concept. Unlike metallic or semiconductor nanoconfiners, voids require no deposition of foreign material and surface functionalization, and no external nanostructure fabrication. The confining element is created directly within the host crystal and can be erased and regenerated post-fabrication. This intrinsic, material-agnostic character distinguishes void-based confinement from all previously demonstrated confiner systems and suggests a route toward momentum-engineered photonics embedded directly within existing device platforms without complex processing steps.

Structural characterization establishes a distinct void-containing phase through a localized melt–quench process, accompanied by W-center formation, dislocation signatures, and a Boson peak consistent with a ~1.3 nm correlation length. These structural fingerprints confirm void formation and lattice reconstruction, but cannot account for the broadband visible-to-near-infrared emission, which is fundamentally incompatible with any known defect luminescence in silicon. *The structural and optical observations are therefore complementary: the former establish the void-containing phase, the latter - its photonic consequence*.

The reversibility of the optical response further distinguishes this phenomenon from conventional defect emitters. The luminescent state can be repeatedly created, erased, and regenerated through alternating electrical conditioning and optical recrystallization, with full recovery of the photoluminescence intensity and spectral profile across multiple write-erase cycles. These observations indicate that the melt–quench process reliably regenerates the void-containing phase and confirm that the optical functionality arises from reversible structural modification of the host lattice rather than permanent chemical transformation.

The demonstration of broadband electroluminescence establishes that momentum-engineered transitions remain accessible under electrical carrier injection. This confirms that nanometric dielectric discontinuities activate broadband emission pathways under both optical and electrical excitation.

More broadly, these findings suggest that the electromagnetic boundary condition imposed by a nanometric dielectric discontinuity - rather than the composition of the confining structure - is the primary determinant of optical functionality in this regime. Silicon represents a particularly stringent test case owing to its large momentum mismatch and indirect band structure, yet the underlying principle is not specific to silicon. Any material system in which optical transitions are limited by momentum conservation may, in principle, benefit from nanometric dielectric discontinuities capable of broadening the accessible photon momentum distribution. This perspective extends the role of optical antennas beyond field enhancement alone, establishing dielectric confinement as a means of engineering momentum space itself. In this picture, optical functionality becomes a property of dielectric confinement rather than material composition, with nanometric dielectric discontinuities providing a new degree of freedom for engineering light–matter interactions at the nanoscale.

## Conflict of interest

The authors declare no competing interests.

## Data availability

The data is available from the corresponding author on reasonable request.

## Acknowledgement

D.A.F. thanks Yulia Davydova for support and help during this study. The authors thank Prof. Joe Patterson, Prof. Maxx Arguilla, Prof. V Ara Apkarian, Dr. Toshihiro Aoki, Pushp Raj Prasad and Prof. Craig Murray for fruitful discussions. Authors thank Amanda Durkin and Mihaela Balu for help with photoluminescence lifetime made available by NIH S10-OD028698 award. D.A.F. and E.O.P acknowledge funding from Chan Zuckerberg Initiative 2023-321174 (5022) GB-1585590, NSF 2434622, and NIH S10-OD028698. D.A.F. dedicates this work to beloved friend Fiona.